\documentstyle[prl,aps,epsfig,twocolumn]{revtex}
\begin{document}
\pagestyle{empty}
\narrowtext
\noindent
{\bf Universal Power-law Decay in Hamiltonian Systems?}
 
\begin{figure}
\begin{center}
\epsfxsize=8.4cm
\leavevmode
\epsffile{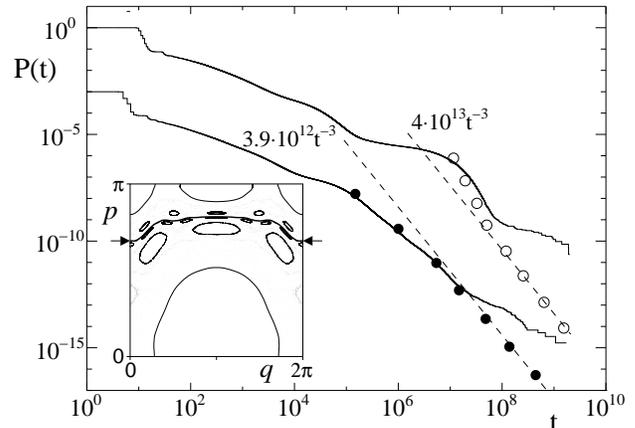}
\caption[fig:tcube]{
The Poincar\'e recurrence time distribution $P(t)$ for the standard map 
at $K=K_c$ for trajectories approaching the critical golden torus from
above (upper curve) and from below (lower curve, shifted by $10^{-3}$).
For large times we find clear deviations from the predictions of 
Ref.~\cite{CS99} (dashed lines and symbols).
Inset: Phase space of the symmetrized standard map at $K=K_c$, with arrows 
pointing at the critical golden torus.
}
\label{fig:tcube}
\end{center}
\end{figure}

\vspace{.3cm}
The understanding of the asymptotic decay of correlations and 
of the distribution of Poincar\'e recurrence times $P(t)$ has been a major challenge
in the field of Hamiltonian chaos for more than two decades.
In a recent Letter, Chirikov and Shepelyansky \cite{CS99} claimed 
the universal decay $P(t) \sim t^{-3}$ for Hamiltonian systems.
Their reasoning is based on renormalization arguments
and numerical findings for the sticking of chaotic trajectories
near a critical golden torus in the standard map.
We performed extensive numerics and find clear deviations from the
predicted asymptotic exponent of the decay of $P(t)$.
We thereby demonstrate that even in the supposedly simple case, when 
a critical golden torus is present, the fundamental question of asymptotic
statistics in Hamiltonian systems remains unsolved.
As in Ref.~\cite{CS99} we study the standard map
\begin{equation}\label{standardmap}
q_{n+1}=q_n+p_n \,{\rm mod}\,2\pi\qquad p_{n+1}=p_n+K\sin q_{n+1} \,\, ,
\end{equation}
at $K=K_c=0.97163540631$, where the golden torus is critical (Fig.~1, inset).
We determine the Poincar\'e recurrence time distributions $P(t)$ 
for trajectories starting below and above the critical golden torus
by using the same numerical approach as in Ref.~\cite{CS99}.
By considerably increasing the statistics we are able to
extend the distribution by almost two orders of magnitude
in recurrence times.
We verify that our statistical data are not affected by the unavoidable
finite numerical precision by comparing data for double ($\approx$16 significant
digits) and quadruple ($\approx$32 digits) precision.
The data for approaching the critical golden torus from above and below
are presented in Fig.~\ref{fig:tcube}.
For times $t < 10^8$ our data agree with the results
presented in Fig.~2 of Ref.~\cite{CS99}.
For larger times, however, we find strong deviations from the predicted
universal power law $P(t) \sim t^{-3}$ (dashed lines in Fig.~\ref{fig:tcube}).
The deviations might be explained in two ways:
The onset of the claimed asymptotic decay might occur for larger times,
which is in contradiction to the prefactors determined in Ref.~\cite{CS99}.
On the other hand, the long-time trapping of chaotic trajectories might be dominated
by islands of stability (non-principal resonances) that are neglected by the
renormalization arguments.  
In fact, the latter possibility is supported by a detailed investigation~\cite{WHK}.
If even in the supposedly simple case of a critical golden torus the
decay $P(t) \sim t^{-3}$ is not observed, the claim for a universal
existence of this decay cannot be maintained.
It thus remains a fundamental challenge in the field of Hamiltonian chaos
whether the asymptotic behavior of $P(t)$ follows a universal power
law and what the value of its exponent would be.

We thank Dima Shepelyansky for candid discussions.

\vspace{.5cm}
\noindent
M.~Weiss$^{1,2}$, L.~Hufnagel$^1$, and R.~Ketzmerick$^1$

\noindent
{\small $^1$ Max-Planck-Institut f\"ur Str\"omungsforschung
and Institut f\"ur Nichtlineare Dynamik der Universit\"at G\"ottingen, Germany}

\noindent
{\small $^2$ EMBL, Meyerhofstr.~1, 69117 Heidelberg, Germany}

\vspace{.5cm}


\vspace{-.5cm}

\begin{references}
\vspace{-1.5cm}

\bibitem{CS99} B.~V.~Chirikov and D.~L.~Shepelyansky, Phys.~Rev.~Lett. {\bf 82}, 
528 (1999).

\bibitem{WHK} M.~Weiss, L.~Hufnagel, and R.~Ketzmerick, nlin.CD/0106021.
\end{references}
\end{document}